# Asset-Class Specific Sustainability Disclosure: Lessons Learned from the EU MiCA Regulation

*Working paper*

*Ulrich Gallersdörfer[1], Lena Klaaßen[1], Christoph Kreiterling[2], Christian Stoll[1]*

## Abstract

Sustainability disclosure in the European Union has so far focused on corporate reporting and sustainability information attached to traditional financial products. Crypto-assets, with consensus-mechanism-driven environmental externalities and fragmented issuer structures, largely fell through this disclosure architecture. The EU Markets in Crypto-Assets Regulation (MiCA) introduces the first EU-wide, asset-class-specific sustainability disclosure regime for crypto-assets by mandating standardised sustainability indicators for both issuers and crypto-asset service providers. Drawing on a policy and legal analysis of MiCA and its Level 2 measures, and on early implementation evidence from public registers and market practice, this paper shows how the policy approach shifted from debates about restricting energy-intensive consensus mechanisms to a transparency regime built on quantitative metrics, machine-readable reporting, and methodological alignment with the broader EU sustainable-finance framework. The paper also highlights practical frictions in implementation, including data gaps, responsibility allocation between issuers and intermediaries, and cross-border supervisory fragmentation. The overview and quantitative summary of the ESMA Interim MiCA Register in this paper may also provide valuable information to regulators and market participants in the context of the European Commission's 2026 targeted consultation on the review of MiCA, including its specific question on environmental and sustainability reporting.

## 1. Introduction

### 1.1. The disclosure gap in the EU sustainable-finance landscape

Sustainable-finance regulation in the European Union has expanded quickly over the past decade. The Sustainable Finance Disclosure Regulation (SFDR) and the Corporate Sustainability Reporting Directive (CSRD) created a broad framework for sustainability-related transparency, but both were designed mainly with traditional financial products and corporate reporting entities in mind (European Parliament and Council 2023).

Crypto-assets sit awkwardly in that architecture. Many token ecosystems are decentralised, many issuers operate across borders or outside classic corporate structures, and the most material environmental issue often sits at network level rather than entity level. That makes it hard to capture climate and other environment-related impacts through frameworks that attach disclosure obligations either to financial market participants and products or to reporting companies. MiCA responds to that structural gap by requiring product-level sustainability information for crypto-assets and related website disclosures by in-scope service providers (European Parliament and Council 2023; European Commission 2025a).

[1] Co-founder of CCRI GmbH, Germany
[2] Trier University of Applied Sciences, Germany

### 1.2. MiCA in context

MiCA's core objective is to create legal certainty, strengthen consumer protection, and harmonise rules for crypto-asset markets across the EU. The Regulation introduces authorisation requirements for crypto-asset service providers (CASPs), conduct rules for issuers, and safeguards aimed at market integrity and financial stability. Sustainability is not the centrepiece of MiCA, but it is built into the regime in a targeted way. Environmental disclosures are required in crypto-asset white papers and, for authorised CASPs, on their websites (European Parliament and Council 2023).

That matters conceptually. MiCA reframes sustainability in crypto markets not as a niche investor preference, but as part of market transparency. In that sense, the Regulation moves the debate away from a binary question of whether high-energy consensus mechanisms should be prohibited and towards a more data-driven model in which market participants, supervisors, and users can compare environmental profiles on a common basis.

## 2. Sustainability disclosure under MiCA

### 2.1. Policy rationale and legislative process

Sustainability was one of the most politically contested parts of the MiCA negotiations. Earlier debates included proposals that would have restricted or effectively phased out energy-intensive proof-of-work models. Those proposals triggered strong pushback from parts of the industry and raised questions about technological neutrality. The final compromise did not impose an outright ban. Instead, it embedded mandatory environmental disclosures and shifted the regulatory approach from prohibition to transparency.

The legislative process involved the European Commission, the European Parliament, and the Council, with technical input from the European Securities and Markets Authority (ESMA) and the European Banking Authority (EBA). Industry actors, NGOs, and researchers also shaped the debate by contesting methods, data availability, and the appropriate level of granularity. The resulting framework is a compromise: it is detailed enough to produce comparable information, but still flexible enough to work with imperfect data and heterogeneous network designs.

### 2.2. Technical standards: content, methodology, and reporting format

The operational framework rests on two distinct Level 2 instruments. First, Commission Delegated Regulation (EU) 2025/422 sets out the sustainability indicators, presentation rules, and methodological principles for the information to be disclosed in white papers and on CASP websites (European Commission 2025a). Second, Commission Implementing Regulation (EU) 2024/2984 specifies the templates and technical format for crypto-asset white papers, including the XHTML and inline XBRL requirements that make the documents machine-readable (European Commission 2024). Keeping those two layers separate is important. The first governs what should be disclosed and how it should be calculated. The second governs how the white paper should be structured and published.

The sustainability indicators themselves are grouped into three categories. The mandatory layer is built around annual energy consumption as the core metric. A supplementary layer adds further indicators, especially for more energy-intensive consensus mechanisms, including renewable-energy share, energy intensity, scope 1 and scope 2 greenhouse-gas emissions, and GHG intensity. A third, optional layer allows additional disclosures, for example on scope 3 emissions, waste electrical and electronic equipment, hazardous waste, and other resource-use metrics (European Commission 2025a).

- Mandatory indicator: annual energy consumption as the baseline metric for all in-scope crypto-assets.
- Supplementary indicators: additional energy and GHG metrics where yearly energy use exceeds 500,000 kWh; for CASP websites, this also depends on the service category.
- Optional indicators: further environmental information such as energy mix, scope 3 emissions, and waste-related metrics.

The Delegated Regulation is deliberately quantitative. It requires gross energy and emissions metrics, does not allow netting through carbon offsets, and permits estimates only where information is not available in a reasonable timeframe and the assumptions are disclosed transparently. It also aligns energy and GHG methodologies with the broader EU sustainability-reporting logic under the CSRD framework (European Commission 2025a).

On the format side, the iXBRL requirement is now operational, not merely prospective. The Implementing Regulation entered into force in December 2024 and has applied since 23 December 2025. ESMA has also published the MiCA White Paper Taxonomy 2025, the reporting manual, validation rules, and Excel-based showcase files to support implementation (European Commission 2024; ESMA 2025; ESMA 2026). That digital layer matters because it turns sustainability disclosure into data infrastructure rather than a purely narrative exercise.

### 2.3. Implementation in practice: register evidence and emerging reporting patterns

Early implementation of MiCA sustainability disclosure is shaped by the interaction of three elements: the sustainability indicator regime, the technical disclosure infrastructure, and the transitional market structure in which authorisation and enforcement still differ across Member States. The move to machine-readable white papers has accelerated the development of standardised templates and third-party tooling. At the same time, public registers now make it possible to observe how the regime is being populated in practice.

ESMA's Interim MiCA Register is the key public source for that state-of-play assessment. It is updated weekly and covers Title II white papers, issuers of ARTs and EMTs, authorised CASPs, and non-compliant entities. The register is valuable, but it also needs to be read carefully. ESMA explicitly notes that the white papers listed in the register have not been reviewed or approved by a competent authority. Register presence therefore signals notification and publication, not substantive endorsement (ESMA 2026).

The figures below summarise the authors' extraction from the Interim MiCA Register as of 3rd July 2026.[3] At this point in time, 282 entities have received a MiCA license as a CASP (see Figure 1). The data also shows that while licenses spread widely across jurisdictions and types of services, there is also a certain level of concentration on a number of jurisdictions and service lines (see Figure 2 and Figure 3). Germany showed the highest number of authorised CASPs in the July snapshot, followed by France, the Netherlands, Malta, and Cyprus. On the service side, authorisations were concentrated in custody and administration, transfer services, exchange services, and execution of orders. By contrast, relatively few firms had obtained authorisation to operate a trading platform for crypto-assets (only 18 entities).

[3] All analyses are based on the Interim MiCA Register. The Interim MiCA Register may contain inaccuracies which we do not correct.

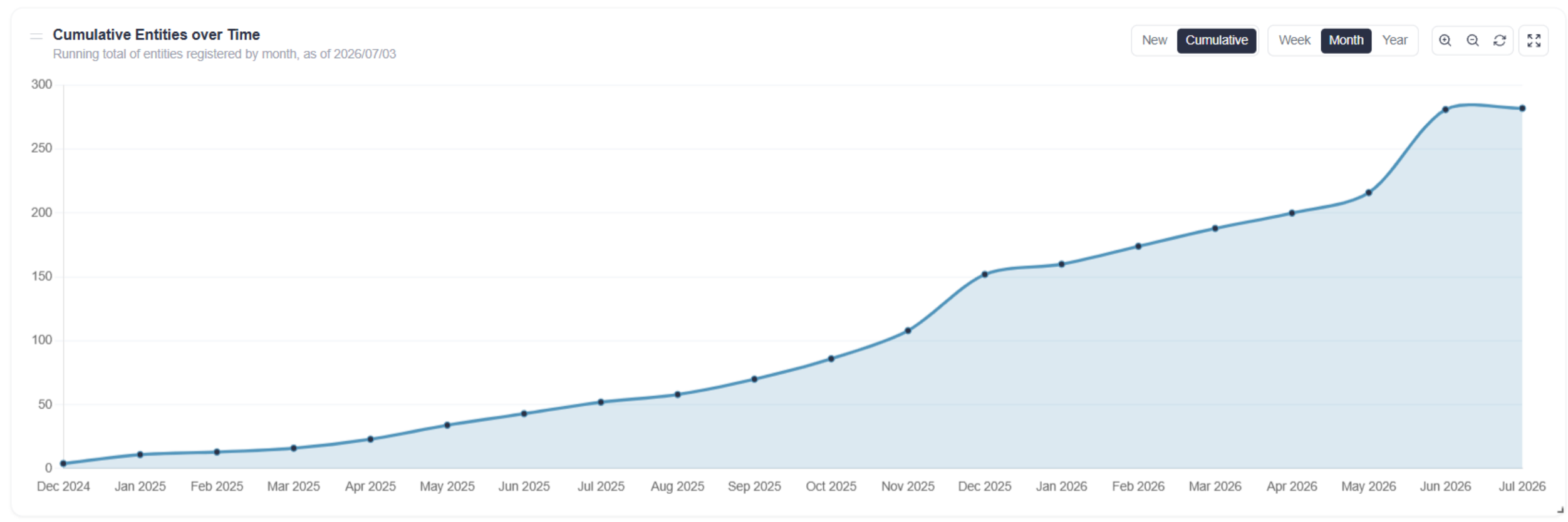

***Figure 1. Overview of cumulative number of entities with CASP MiCA licenses over time as of 3rd July 2026. Source: [https://mica-monitor.com/](https://mica-monitor.com/) based on the Interim MiCA Register***

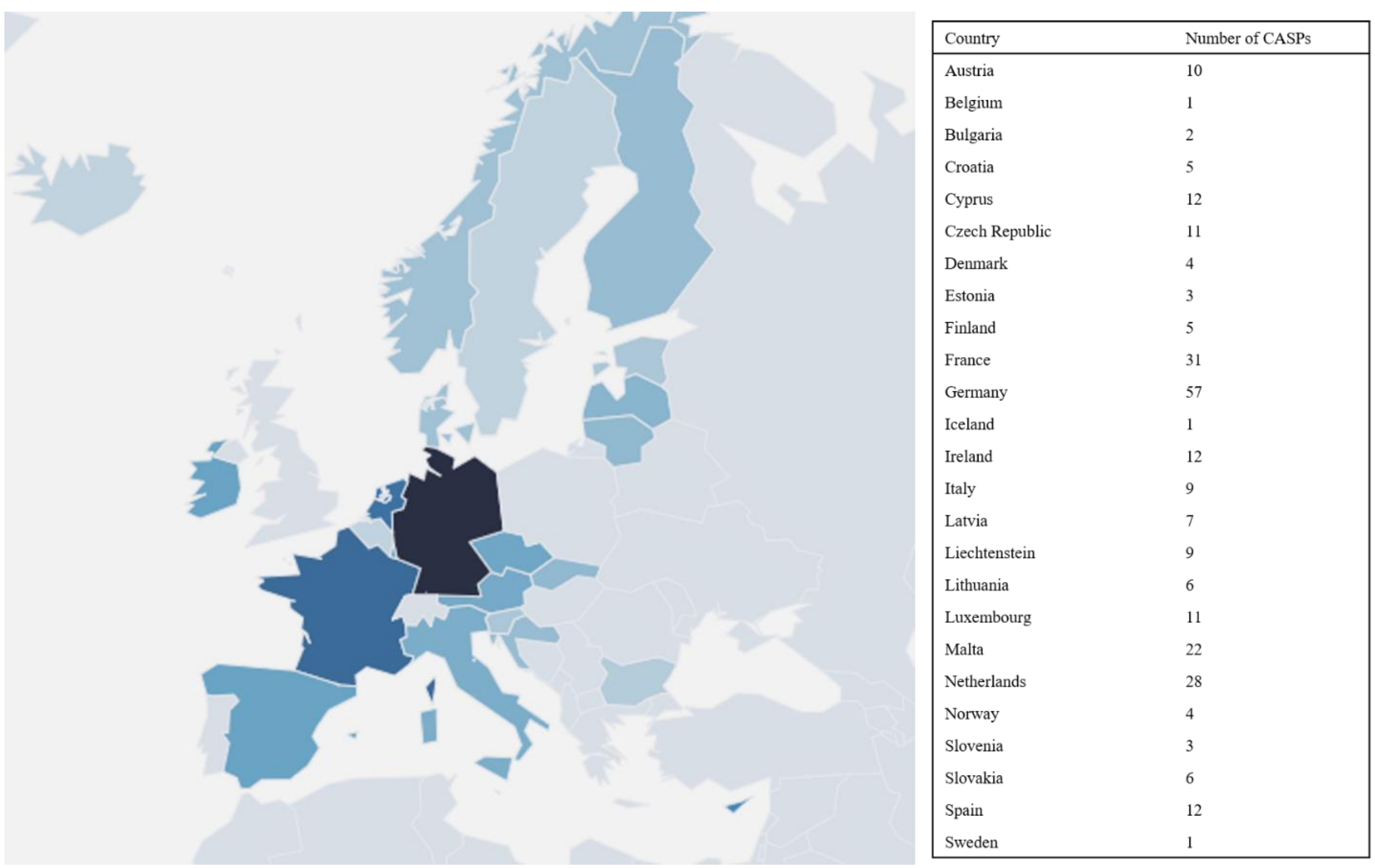

| Country | Number of CASPs |
|---|---|
| Austria | 10 |
| Belgium | 1 |
| Bulgaria | 2 |
| Croatia | 5 |
| Cyprus | 12 |
| Czech Republic | 11 |
| Denmark | 4 |
| Estonia | 3 |
| Finland | 5 |
| France | 31 |
| Germany | 57 |
| Iceland | 1 |
| Ireland | 12 |
| Italy | 9 |
| Latvia | 7 |
| Liechtenstein | 9 |
| Lithuania | 6 |
| Luxembourg | 11 |
| Malta | 22 |
| Netherlands | 28 |
| Norway | 4 |
| Slovenia | 3 |
| Slovakia | 6 |
| Spain | 12 |
| Sweden | 1 |

***Figure 2. Geographical distribution of authorised MiCA CASPs as of 3rd July 2026. Source: [https://mica-monitor.com/](https://mica-monitor.com/) based on the Interim MiCA Register***

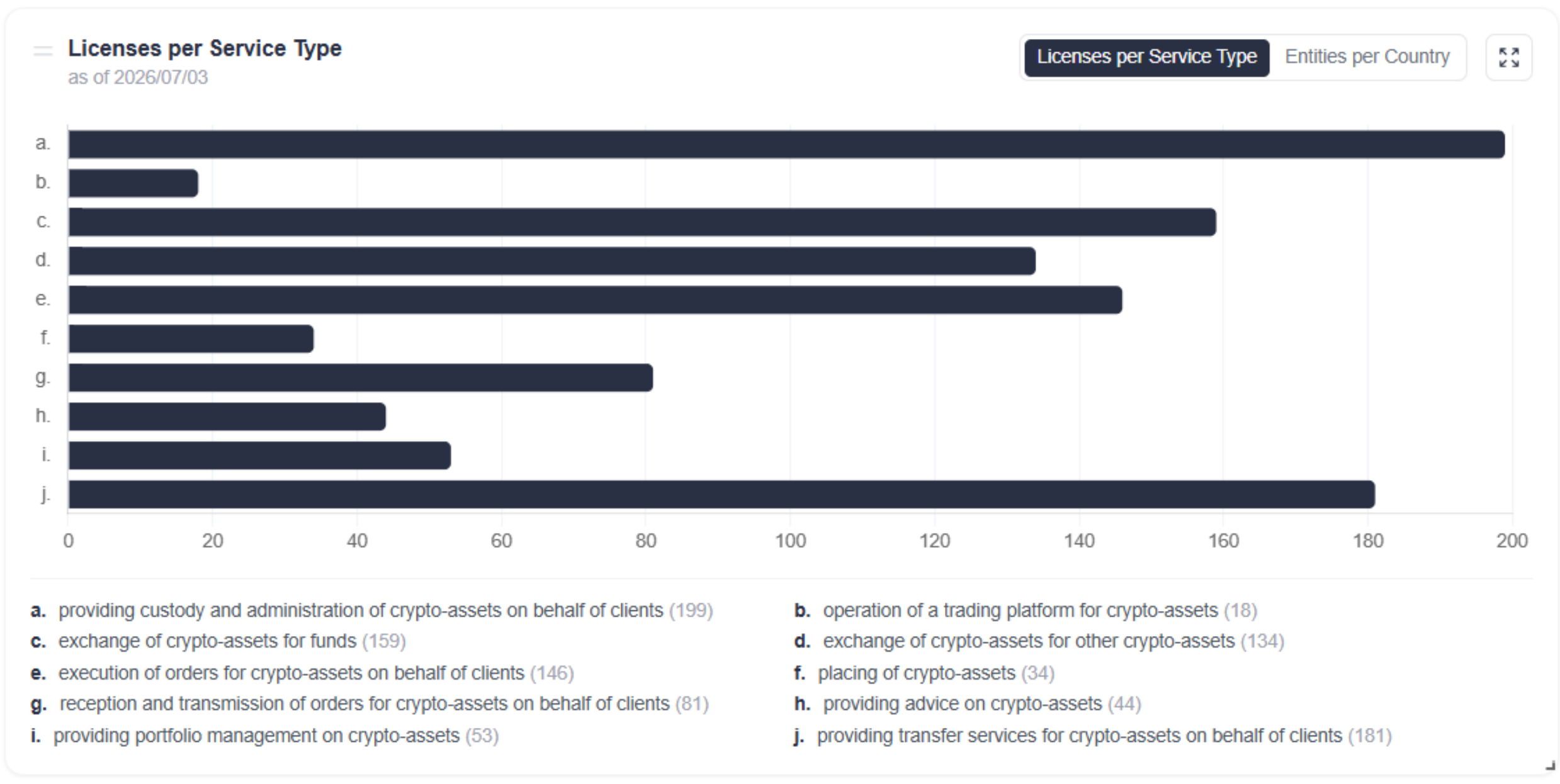

***Figure 3. Service categories for which CASPs were authorised as of 3rd July 2026. Source: https://mica-monitor.com/ based on the Interim MiCA Register***

Regarding the white papers, Figure 4 shows an overview by country of the respective competent authority. For white papers other than EMTs and ARTs, most are notified in Ireland (340), followed by Malta (155) and Germany (141) while for EMTs most are notified in the Netherland (9), followed by France (8) and Germany (4). Another implementation pattern is emerging around the relationship between white-paper disclosure and CASP-level publication. In principle, CASPs should be able to rely on upstream sustainability information contained in white papers. In practice, that chain is often incomplete. White papers vary in granularity, methodology, and update discipline, and decentralised networks may lack a clear central actor responsible for maintaining high-quality environmental data over time. The result is a hybrid implementation model in which formal disclosure obligations still sit with issuers and authorised CASPs, but the operational packaging of sustainability information is increasingly supported by specialist data providers and reporting tools.

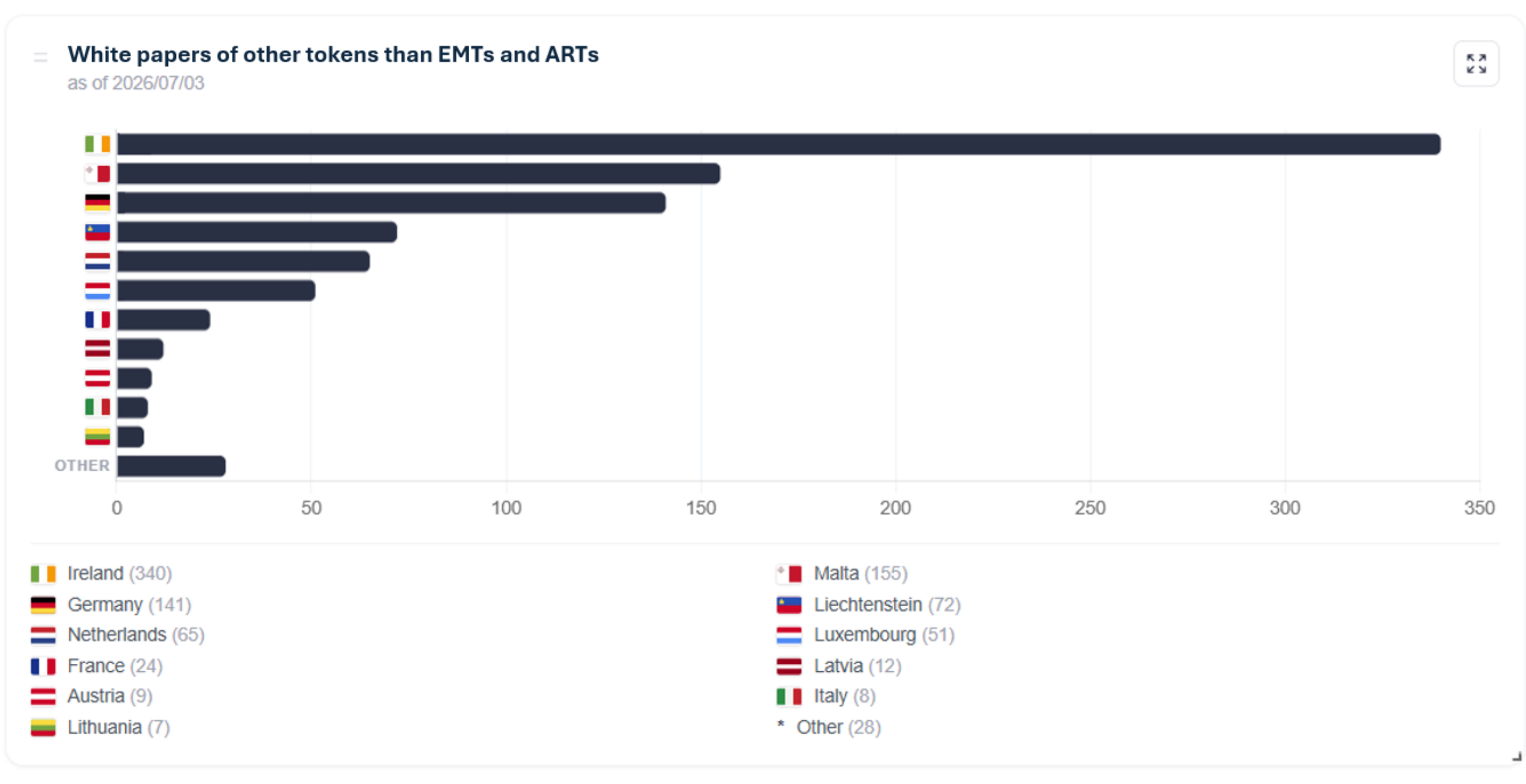

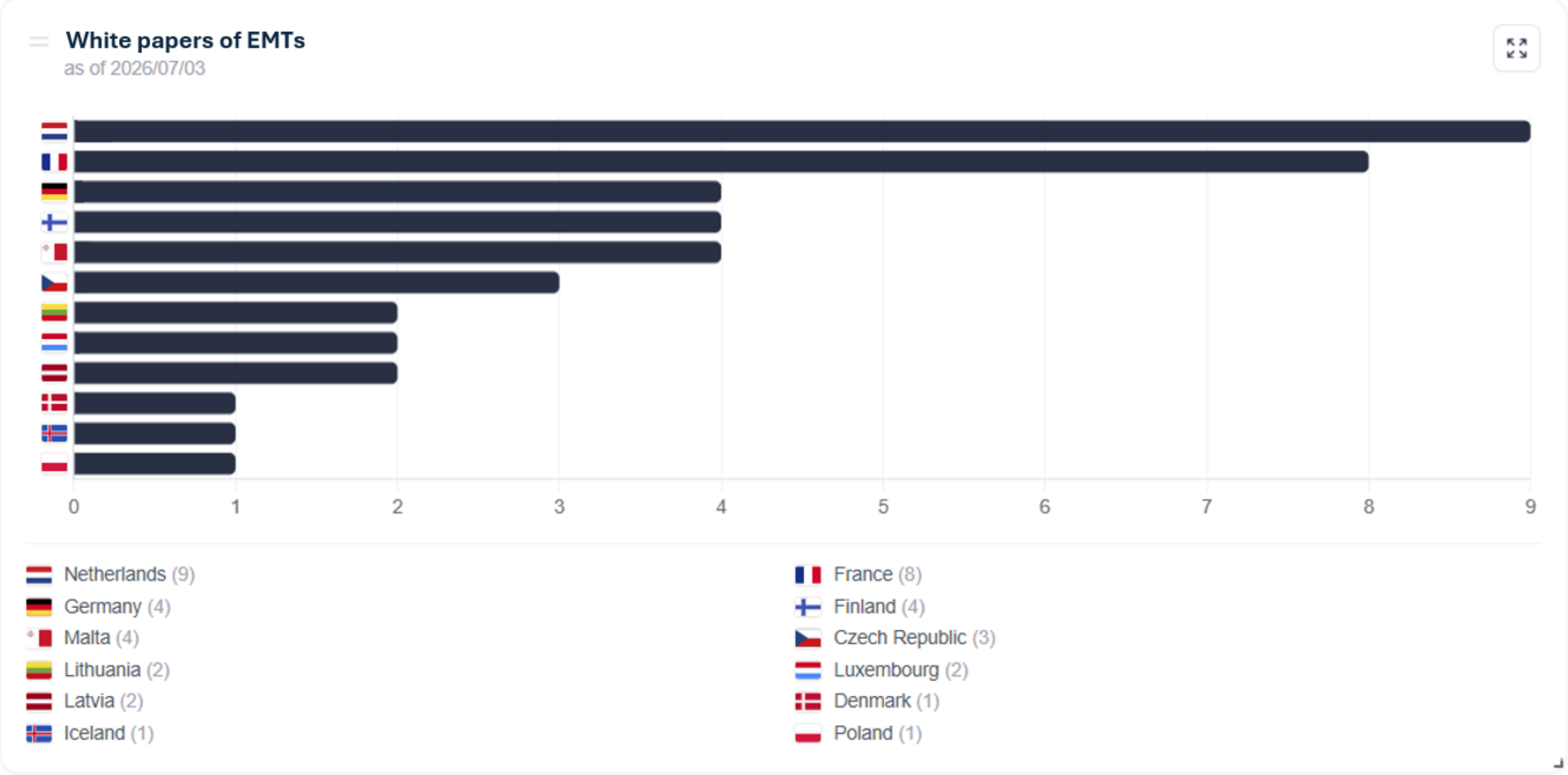

***Figure 4. Overview of white papers by country of the respective competent authority as of 3rd July. The top represents the white papers of other tokens than EMTs and ARTs. The bottom represents white papers of EMTs. As of 3rd July no ART is present in the register. Source: [https://mica-monitor.com/](https://mica-monitor.com/) based on the Interim MiCA Register***

## 3. Insights, frictions, and future outlook

### 3.1. Interoperability with other sustainability reporting frameworks and standards

MiCA's sustainability indicators were not designed in isolation. Their logic is tied to the broader EU sustainable-finance architecture. Three alignment points matter in particular.

- GHG methodologies draw on concepts that are compatible with the GHG Protocol and the ESRS E1 climate standard.

- Energy-consumption metrics create data points that can inform principal-adverse-impact analysis under the SFDR.
- Machine-readable white papers make it easier to integrate crypto-asset disclosures into wider digital reporting pipelines.

This does not create full methodological equivalence across frameworks, but it does reduce duplication. MiCA generates product-level environmental data that may later feed into entity-level or portfolio-level disclosures. In that sense, it is best understood as a complement to SFDR and CSRD rather than a parallel disclosure universe.

### 3.2. Regulatory burden versus added value

The compliance burden is real, especially for intermediaries that do not control the underlying network but still need to publish sustainability information. At the same time, the added value of disclosure should not be underestimated. Better data can reduce information asymmetries, support institutional due diligence, and differentiate firms that can provide credible sustainability information.

The Zumo MiCA Readiness Report offers a useful early snapshot, even if its sample is small and not representative of the whole market. It found that 75% of respondents described themselves as very familiar with MiCA overall, but only 31% said they were knowledgeable about MiCA's sustainability requirements. Another 50% said they were aware of those requirements without describing themselves as knowledgeable. When asked specifically about sustainability compliance readiness, 25% said they were already prepared to comply, 63% were still exploring options, and 13% had not taken any steps. The most frequently reported pain points were unclear regulatory requirements and expectations (50%), lack of ready-made solutions (50%), resource burden (38%), and lack of information (31%) (Zumo 2025).

Those findings support a balanced reading. MiCA sustainability disclosure is not costless, and part of the burden falls on firms that act as intermediaries rather than originators of the data. But the same regime may also accelerate market maturation by rewarding firms that build credible reporting capabilities early.

### 3.3. Supervision, methodological guidance, and assurance

Primary supervisory responsibility for environmental disclosures sits with national competent authorities, with ESMA and the EBA providing coordination and guidance. That structure leaves room for divergence, especially where supervisors must assess the reliability of environmental data for decentralised networks with diffuse governance and incomplete observable inputs.

The transitional phase which ended on 30 June 2026 made that more complicated. Article 143 allowed Member States to apply different grandfathering periods, and ESMA's published list shows that national approaches varied materially. ESMA has also clarified that entities operating under the grandfathering regime are not CASPs within the meaning of MiCA until they are authorised. That distinction mattered for any empirical analysis of implementation and for any claim about who is already fully inside the MiCA disclosure perimeter (ESMA 2024a; ESMA 2024b; ESMA 2024c). As the transitional phase ended on 30 June 2026, only CASPs within the meaning of MiCA may operate now.

A further question is whether the current NCA-led model will remain unchanged. In December 2025, the European Commission presented its market integration and supervision package and explicitly proposed assigning ESMA direct oversight over certain significant market infrastructures and crypto service providers. That proposal is not law yet, but it signals that supervisory centralisation is now part of the policy debate (European Commission 2025b; European Commission 2025c).

Assurance is another open issue. MiCA does not currently impose a dedicated audit requirement for sustainability metrics, but pressure for more formal verification may grow as disclosures become more decision-relevant and more deeply integrated into EU reporting systems.

### 3.4. Future developments in sustainability disclosure

In the near term, the most likely developments are operational rather than conceptual. More market participants will move to structured iXBRL publication. More data providers will build tooling around white-paper extraction, comparison, and website publication. And the EU's broader reporting infrastructure will make those disclosures easier to reuse.

One important reference point is the European Single Access Point (ESAP). ESAP's first data-collection phase began in July 2026, but it does not include MiCA information. The public platform is scheduled to become available by July 2027, while MiCA information is scheduled to enter ESAP from 10 January 2030. ESAP is therefore a longer-term rather than immediate channel for MiCA sustainability disclosures (European Parliament and Council 2023; Joint Committee of the ESAs 2024).

Longer term, policy questions remain open. A later review of MiCA could lead to more granular environmental indicators, stronger expectations around verification, or some extension beyond purely environmental metrics. Whether the regime will eventually connect more directly to taxonomy-style classification or broader sustainability-labelling debates is still an open question, not a settled path.

## 4. Conclusion and research directions

MiCA's sustainability disclosure framework is a significant step in integrating crypto-assets into the European sustainable-finance landscape. Its main contribution is not that it solves the underlying environmental challenges of crypto markets. Its contribution is that it makes those challenges more visible, more comparable, and more actionable. Indeed, even though the quality of the sustainability disclosures in white papers varies, the reporting framework for CASPs appears operationally workable. The regulation's clear structure has created a solid foundation that allowed solutions from specialized data providers and dedicated reporting tools to rapidly emerge, although the register evidence alone does not establish the completeness, comparability, or substantive quality of the disclosures.

Several broader lessons follow. Asset-class-specific disclosure can make sense when the underlying technology creates environmental impacts that are poorly captured by entity-level or product-level frameworks designed for other markets. Standardised, machine-readable data can reduce information asymmetry and support both supervision and market use. But implementation also shows that disclosure regimes work only if data responsibilities, update processes, and supervisory expectations are clear.

That leaves a clear research agenda. Future work could examine the quality and comparability of sustainability data across crypto-asset white papers, the extent to which CASPs rely on issuers versus third-party providers, the behavioural effect of sustainability labels on users and investors, and the conditions under which assurance might become necessary. MiCA is therefore best seen as an early test bed: not the final model for crypto sustainability disclosure, but an important first iteration.